\documentclass{aa}  
\usepackage{graphicx}
\usepackage{txfonts}
\def\kms  {km~s$^{-1}$}

\def\Vlsr {\ifmmode {V_{\rm LSR}} \else {$V_{\rm LSR}$} \fi}
\def\Ro   {\ifmmode {R_0} \else {$R_0$} \fi}
\def\To   {\ifmmode {\Theta_0} \else {$\Theta_0$} \fi}

\begin{document}
   \title{The nature of the methanol maser ring G23.657$-$00.127}

   \subtitle{I.\,The distance through trigonometric parallax measurements}

   \author{A. Bartkiewicz
          \inst{1},
          A. Brunthaler
          \inst{2},
          M. Szymczak
          \inst{1},
          H.J. van Langevelde\inst{3,4},
          \and
          M.J. Reid\inst{5}
          }

%   \offprints{A. Bartkiewicz}

   \institute{Toru\'n Centre for Astronomy, Nicolaus Copernicus
          University, Gagarina 11, 87-100 Toru\'n, Poland\\
          \email{[annan;msz]@astro.uni.torun.pl}
\and      Max-Planck-Insitut f\"ur Radioastronomie, Auf dem H\"ugel 69, 53121 Bonn, Germany\\
          \email{brunthal@mpifr-bonn.mpg.de}
\and      Joint Institute for VLBI in Europe, Postbus 2, 7990 AA
          Dwingeloo, The Netherlands\\
          \email{langevelde@jive.nl}
\and      Sterrewacht Leiden, Leiden University, Postbus 9513, 2300 RA Leiden, The Netherlands
\and      Harvard-Smithsonian Center for Astrophysics, 60 Garden Street, Cambridge, MA 02138, USA\\
          \email{reid@cfa.harvard.edu}
          }

   \date{Received 26 June 2008/Accepted 10 September 2008}

\authorrunning{Bartkiewicz et al.}
\titlerunning{The nature of the methanol maser ring}

% \abstract{}{}{}{}{} 
% 5 {} token are mandatory
 
  \abstract
   {Methanol masers are 
associated with young high-mass stars and are an important 
tool for investigating the process of massive star formation.}
   {The recently discovered methanol maser ring in G23.657$-$00.127 provides 
an excellent ``laboratory'' for a detailed study of the nature and physical 
origin of methanol maser emission, as well as parallax and proper motion 
measurements.}
   {Multi-epoch observations of the 12.2\,GHz methanol maser line from the 
ring were conducted using the Very Long Baseline Array. Interferometric
observations with milliarcsecond resolution enabled us to track single
maser spots in great detail over a period of 2 years.}
   {We have determined the trigonometric parallax 
of G23.657$-$00.127 to be 0.313$\pm$0.039\,mas, giving a distance of
$3.19^{+0.46}_{-0.35}$\,kpc.  The proper motion of the source indicates 
that it is moving with the same circular velocity as the LSR, but it 
shows a large peculiar motion of $\approx35$ km\,s$^{-1}$ toward the Galactic 
center.}
   {}


   \maketitle
%
%________________________________________________________________

\section{Introduction}
Two spectral lines of methanol masers, 6.7\,GHz and 12.2\,GHz, are strongly 
associated with young high--mass stars or protostars and enable one to 
investigate the stellar environment on scales of 100--1000\,AU
in great detail (Menten \cite{menten}; Moscadelli et al. \cite{moscadelli}; 
Minier et al. \cite{minier00}). 
Interferometric studies of methanol masers have been carried out
for more than ten years in order to constrain {\it how} and {\it where}
massive stars are being born (Norris et al. \cite{norris}; Phillips et al.
\cite{phillips}; Walsh et al. \cite{walsh98}; Dodson et al. \cite{dodson}). 
We started interferometric studies of the 6.7\,GHz masers discovered in 
the Toru\'n unbiased survey towards the Galactic plane 
(Szymczak et al. \cite{szymczak02}). 

One source, G23.657$-$00.127, was imaged during the European VLBI Network
(EVN) session in 2004 when eight antennas 
were working for the first time at 5\,cm wavelength. These observations  
revealed a new class of spherically symmetric methanol maser sources
(Bartkiewicz et al. \cite{bartkiewicz}). The ``ring'' of masers in
G23.657$-$00.127 has a mean radius of 127\,mas. Its circular geometry makes
it a unique laboratory for detailed study of a methanol maser in an 
isolated massive star forming region. However, the morphology and 
velocity signature of the maser spots at 6.7\,GHz did not allow us to 
determine unambiguously the origin of the ring--like structure. Therefore we 
started VLBI observations of the 12.2\,GHz methanol maser line following 
its detection with the Toru\'n 32\,m antenna. 

In this paper we report multi--epoch observations of G23.657$-$00.127 
at 12.2\,GHz using the NRAO\footnote{The National Radio Astronomy 
Observatory is operated by Associated Universities, Inc., under a 
cooperative agreement with the National Science Foundation.} 
Very Long Baseline Array (VLBA). We focus on the measurement of the source 
parallax and proper motion, providing a reliable distance. 
Maser variability and internal proper motion studies for both the
6.7 and 12.2\,GHz lines will be presented in a separate paper. 

\section{Observations and data analysis}
\subsection{Observations}
Our first interferometric observations of the methanol maser in 
G23.657$-$00.127 at 12178.597\,MHz using the NRAO Very Large Baseline Array 
(VLBA) were carried out on 2005 November 28 (epoch 1). After successfuly detecting
and imaging the maser we made five additional observations between 2006 June 10 and 
2007 June 19 (see details in Table \ref{table:1}). 
Ten antennas were used during all sessions but only nine antennas 
were working in epoch 3, as Hancock was not available then.
Each observation lasted 10 hours and we used one 8\,MHz band in dual 
circular polarization mode. The band was centered at the Local Standard of Rest 
(LSR) velocity of 83\,km\,s$^{-1}$, which corresponded to the brightest feature of 
6.7\,GHz emission (Bartkiewicz et al. \cite{bartkiewicz}).

We used the VLBA calibrator J1825--0737 (Fomalont 
et al. \cite{fomalont}) as the phase-reference source. The flux densities 
of the phase calibrator in all observing sessions are summarized in Table 
\ref{table:1}.  The pointing position for G23.657$-$00.127
was chosen to be near the brightest spots detected at 6.7\,GHz: 
RA$=$18$^{\rm h}$34$^{\rm m}$51\fs565, 
Dec=$-$08\degr18\arcmin21\farcs3045 (J2000). The separation between 
the phase-reference and maser source was $\sim$2\fdg4. 
For epoch 1 we used a 2.5\,min switch cycle between the phase 
reference source (slewing plus on-source time = 1\,min) and the maser source
(slewing plus on-source time = 1.5\,min).
The total on-source time for the maser was 
5\,hours. In addition, we observed the strong source, J1824$+$0119 (Petrov et 
al. \cite{petrov}), every hour for 2\,min in order to improve the delay 
calibration. 3C454.3 was used as a fringe finder (Ma et al. \cite{ma}). 

Starting with the second epoch, we shortened the time devoted to the
maser source from 1.5 to 1.0\,min, in order to obtain
better phase transfer. We also included a second extragalactic background 
source, J1833$-$0855 (Xu et al.\cite{xu}), alternating with
16\,min cycles of rapid switching between J1825--0737 and J1833--0855
and between J1825--0737 and G23.657$-$00.127. Unfortunately,
J1833--0855 was not useful for astrometry, since the source was
weak, partially resolved and not detected in all epochs.  
The total on-source time for the maser was 2.3\,hours
for epochs 2 -- 6. 

Starting with epoch 2, we also performed {\it geodetic--like} observations 
using 39 ICRF quasars whose positions were known to better than 1\,mas. 
Typically 17 quasars were observed over a timespan of 40\,min, with a frequency setup 
involving eight 4\,MHz bands at left circular polarization that spanned a 
frequency range of 470\,MHz. Three of these blocks were placed 
at the beginning, in middle and end of these 
five sessions. This enabled us to estimate tropospheric zenith delay errors 
and clock offsets from the correlator model (see Reid \& Brunthaler \cite{reid}
and Brunthaler et al. \cite{brunthaler} for a detailed discussion).

The data were correlated at the VLBA correlator in Socorro, NM. 
The phase-referencing data were correlated with 1024 spectral channels,  
corresponding to a channel spacing of 0.193\,km\,s$^{-1}$. The geodetic 
block data, containing the observations of the ICRF quasars, were 
correlated with 16 spectral channels.

\begin{table}
\caption{Details of the VLBA observations}             
\label{table:1}      
\centering          
\begin{tabular}{lcccccc}     
\hline\hline       
Epoch &Date & J1825-0737 & Beam& Image rms\\
      &      & Flux density  & HPBW; P.A.&per channel\\ 
      &      & (mJy) & (mas$^2$); (\degr) & (mJy\,beam$^{-1}$) \\
\hline                    
\\
   1 & 2005 Nov 28 & 166 &3.7$\times$1.2; $-$13& 10\\
   2 & 2006 Jun 10 & 204 &2.3$\times$1.1; $-$5 &7 \\  
   3 & 2006 Sep 21 & 208 &2.2$\times$1.6; $-$3 &7 \\
   4 & 2006 Dec 14 & 219 &2.5$\times$0.9; $-$12&7 \\
   5 & 2007 Mar 21 & 278 &2.2$\times$1.3; $+$1 &7 \\
   6 & 2007 Jun 19 & 275 &2.3$\times$1.1; $-$6 &7 \\
\\
\hline\hline                  
\end{tabular}
\end{table}

\subsection{Calibration and imaging}                      
Most of the calibration and data reduction were carried out with standard
procedures for spectra--line observations using the Astronomical Image
Processing System (AIPS). First, we applied the latest values of the Earth's 
orientation parameters and corrected for effects of the changing 
antenna parallactic angles.  We removed atmospheric zenith delay errors 
using the {\it geodetic--like} observations of the quasars. 
Then we calibrated the visibility amplitudes 
using measured antenna gains and system temperatures and corrected for 
voltage offset in the samplers. Next, we corrected for electronic 
delay and phase differences among the IF bands using data from J1824$+$0199
and applied the phase--referencing technique using J1825$-$0737. 
Finally the interferometer spectra for G23.657$-$00.127 were shifted 
to keep the LSR velocity of 83 \kms\ centered in
the band. Because, in this paper, we are concentrating on the astrometry, 
we do not consider here the results of self--calibration of the maser. 
A detailed disscussion of nature of the 12.2\,GHz emission towards 
G23.657$-$00.127 will be presented in a forthcoming paper.

\begin{figure*}
\centering
\includegraphics[width=16cm]{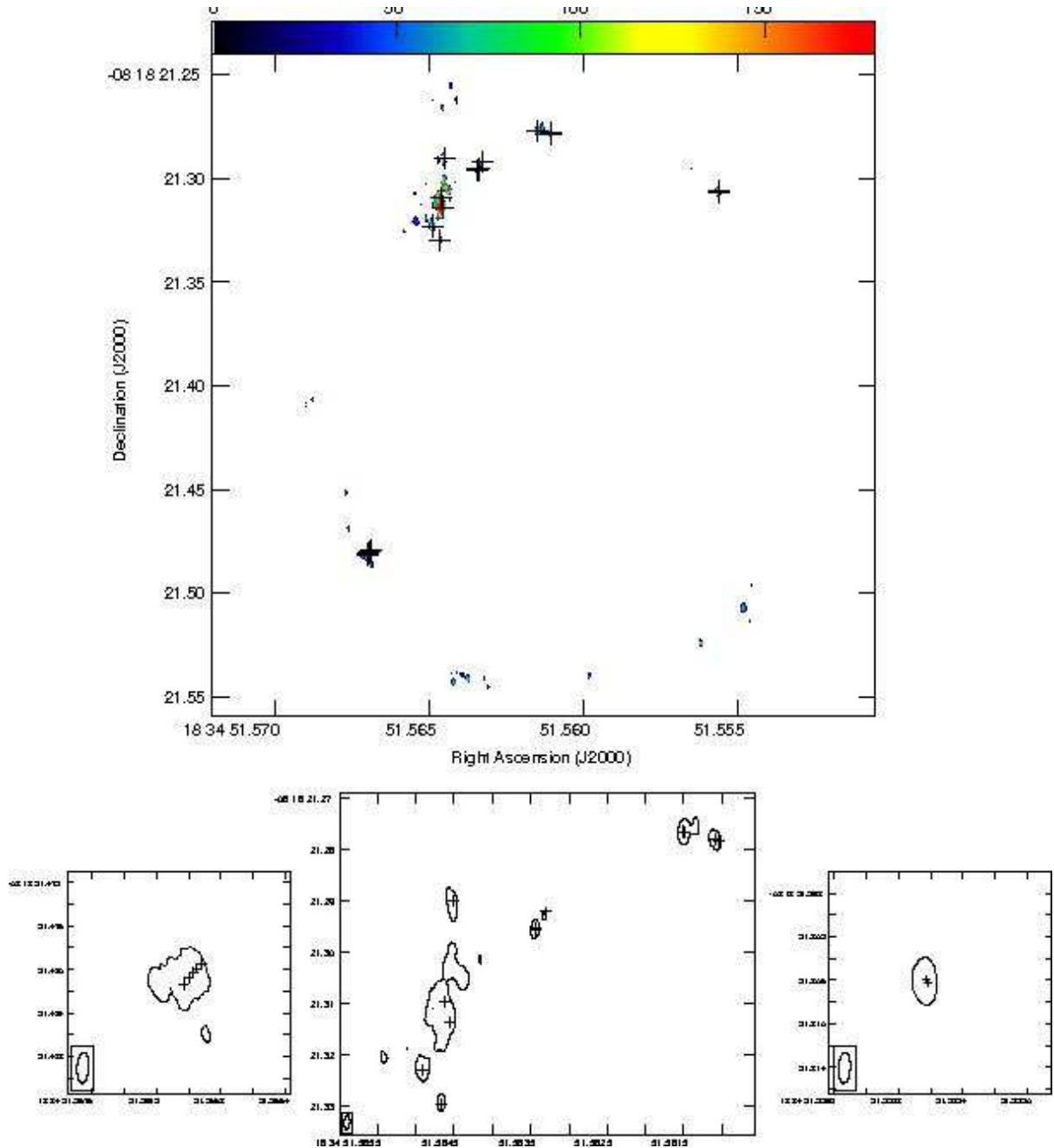}
\caption{{\bf Top:} Total intensity (``zeroth'' moment) map of 12.2\,GHz methanol maser
from G23.657$-$00.127 from 2007 June 19 using the VLBA. The color scale 
(in the electronic version) varies linearly from 0 to 
180\,Jy\,beam$^{-1}$\,m\,s$^{-1}$. The 19 crosses represent spots that were chosen 
for the parallax estimation (Table \ref{table:2}).
{\bf Bottom:} The enlarged views of the masing regions. The contours
correspond to the value of 10\% of the peak (i.e.,
38.6\,Jy\,beam$^{-1}$\,m\,s$^{-1}$). The beam is indicated by the ellipses in
the bottom left--hand corners. 
             }
\label{fig1}
\end{figure*}

\begin{figure*}
\centering
\includegraphics[angle=-90,width=12cm]{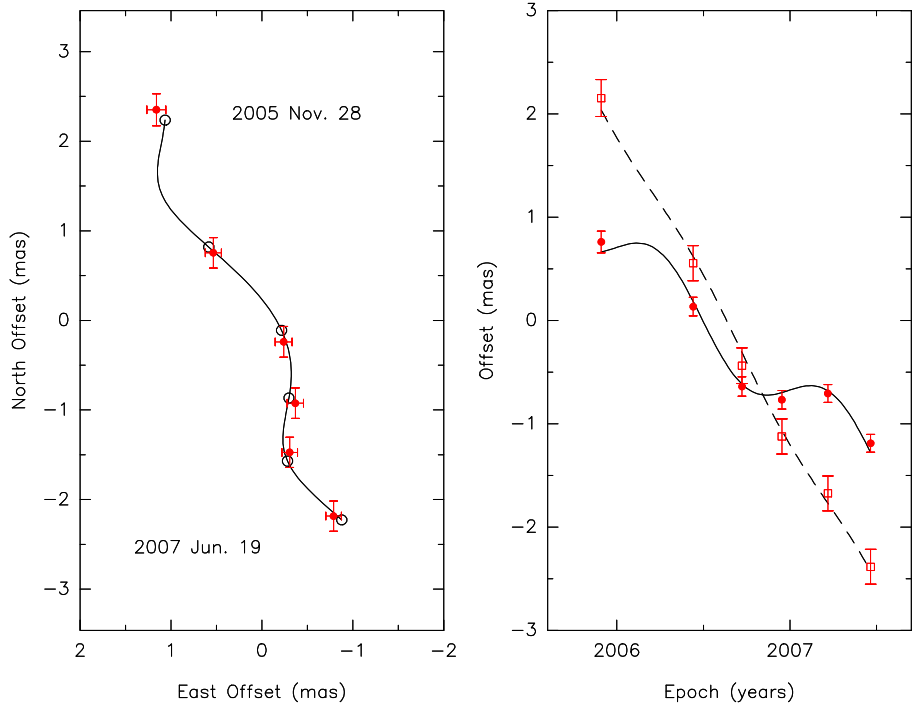}
\caption{{\bf Left:} Six epoch positions of spot 19  (from Table~\ref{table:2}) 
on the plane of the sky with the best parallax and proper motion fit 
superposed. Open circles indicate the location of the model value at the 
dates of observation.
{\bf Right:} Position of spot 19 versus time with the best parallax
and proper motion fit in right ascension (filled circles, solid line) 
and declination (squares, dashed line). 
             }
\label{fig2}
\end{figure*}
\begin{figure*}
\centering
\includegraphics[angle=0,width=10cm]{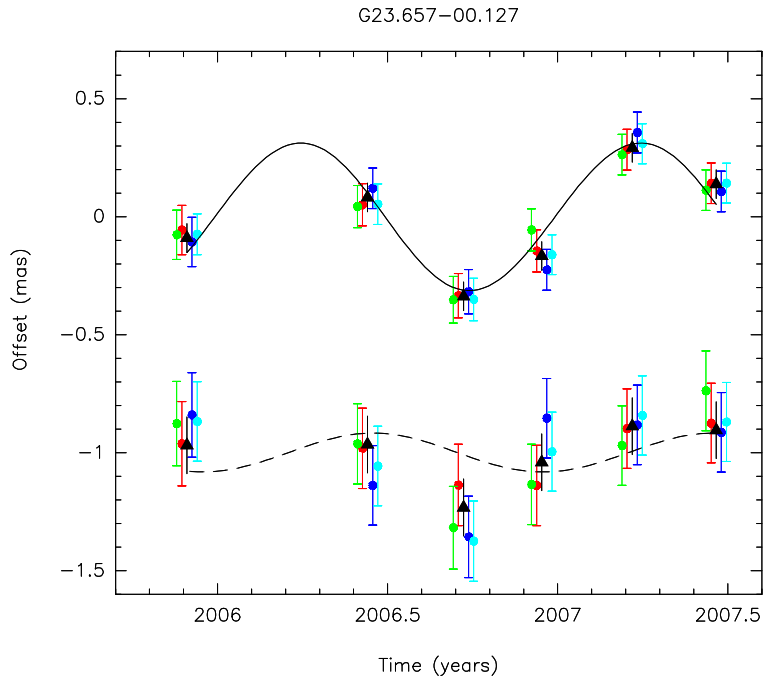}
\caption{Position offsets of four spots (filled circles) in right 
ascension (top) and declination (bottom) after removing the proper motion,
superposed on the combined parallax fit model for all maser spots in 
Table~\ref{table:2}. The data points are slighlty shifted in time for clarity. 
Also shown are the average positions of all spots (black triangles).
             }
\label{fig3}
\end{figure*}

\begin{table*}
\caption{Spots which appeared at all six epochs. Presented coordinates
(columns three and four) are relative to the phase center 
(RA$=$18$^{\rm h}$34$^{\rm m}$51\fs565, Dec=$-$08\degr18\arcmin21\farcs3045) 
at epoch 2. Column five lists the parallax estimates. Columns six and
seven give the motion on the plane of the sky along the right ascension and
declination, respectively.  $\mu_{RA}$ is the true RA motion in
mas\,yr$^{-1}$ and includes the 15\,cos(Dec) factor.
}  
\label{table:2}      
\centering          
\begin{tabular}{llrrccc}     
\hline\hline       
Spot &V$_{\rm LSR}$ & $\Delta$RA & $\Delta$Dec& Parallax &\multicolumn{2}{c}{Proper motion}\\
     &              &           &            &$\pi$   & $\mu_{\rm RA}$ & $ \mu_{\rm Dec}$ \\
     &(km\,s$^{-1}$)&  (mas) & (mas)& (mas)   & (mas\,yr$^{-1}$) & (mas\,yr$^{-1}$) \\
\hline     
\\
1 & 86.837 & $-$38.090 & 16.134  & 0.299$\pm$0.071 & $-$1.20$\pm$0.10 & $-$2.90$\pm$0.16 \\
2 & 86.452 & $-$36.343 & 13.110  & 0.330$\pm$0.100 & $-$1.28$\pm$0.13 & $-$2.87$\pm$0.21 \\
3 & 86.260 & $-$36.149 & 12.916  & 0.306$\pm$0.095 & $-$1.26$\pm$0.12 & $-$2.99$\pm$0.16 \\
4 & 84.335 & $-$152.281&  2.371  & 0.332$\pm$0.070 & $-$1.45$\pm$0.09 & $-$3.12$\pm$0.12 \\ 
5 & 84.143 & $-$152.105& 2.719   & 0.273$\pm$0.057 & $-$1.37$\pm$0.07 & $-$3.07$\pm$0.10 \\
6 & 84.143 & $-$71.635 & 30.214  & 0.269$\pm$0.058 & $-$1.25$\pm$0.07 & $-$2.99$\pm$0.12 \\
7 & 83.950 & $-$17.850 &$-$20.833& 0.336$\pm$0.054 & $-$1.35$\pm$0.06 & $-$3.07$\pm$0.09 \\
8 & 83.950 & $-$71.029 & 30.442  & 0.245$\pm$0.052 & $-$1.43$\pm$0.07 & $-$3.02$\pm$0.11 \\
9& 83.373 & $-$65.264 & 31.895  & 0.373$\pm$0.070 & $-$1.19$\pm$0.10 & $-$3.00$\pm$0.16 \\ 
10& 83.180 & $-$65.064 & 32.004  & 0.311$\pm$0.070 & $-$1.32$\pm$0.08 & $-$3.03$\pm$0.13 \\
11& 82.988 & $-$19.591 & $-$5.390& 0.290$\pm$0.064 & $-$1.37$\pm$0.08 & $-$2.81$\pm$0.11 \\
12& 82.603 & $-$14.350 & $-$14.133&0.277$\pm$0.099 & $-$1.33$\pm$0.13 & $-$3.15$\pm$0.17 \\
13& 82.603 & $-$18.516 & $-$1.136& 0.363$\pm$0.085 & $-$1.30$\pm$0.11 & $-$2.75$\pm$0.15 \\ 
14 & 81.833 & $-$20.069 & 18.417  & 0.301$\pm$0.056 & $-$1.27$\pm$0.07 & $-$2.87$\pm$0.11 \\
15& 80.486 & 15.676  & $-$171.016& 0.406$\pm$0.096 & $-$1.43$\pm$0.11 & $-$2.87$\pm$0.19 \\
16& 80.294 & 16.118 &  $-$171.487& 0.355$\pm$0.099 & $-$1.35$\pm$0.13 & $-$2.96$\pm$0.18 \\
17& 80.101 & 16.429 &  $-$171.892& 0.311$\pm$0.074 & $-$1.35$\pm$0.10 & $-$2.91$\pm$0.14 \\
18& 79.909 & 16.871 & $-$172.332 & 0.249$\pm$0.065 & $-$1.26$\pm$0.08 & $-$2.85$\pm$0.12 \\
19& 79.716 & 17.326 & $-$172.979 & 0.304$\pm$0.053 & $-$1.38$\pm$0.07 & $-$2.92$\pm$0.11 \\
\hline
\\
\multicolumn{2}{l}{Averaging fits} & & & 0.312$\pm$0.010  & $-$1.32$\pm$0.02  &$-$2.96$\pm$0.03 \\
\multicolumn{2}{l}{Combined fit} & & & 0.313$\pm$0.015\\
\multicolumn{2}{l}{Averaging data} & & & 0.313$\pm$0.039\\
\\
\hline\hline                  
\end{tabular}
\end{table*}

We imaged an area of 0.4$\times$0.4\,arcsec$^2$ with four maps, 
each covering (0.2$\times$0.2\,arcsec$^2$).  Each map contained
1024$\times$1024 pixels of size 0.2\,mas and natural weighting was 
applied during mapping.  We searched for emission over the
\Vlsr\ range from 72 to 92\,km\,s$^{-1}$.   The resulting 
beam sizes and rms noise levels (1$\sigma$) in line--free channels for each 
session are summarized in Table \ref{table:1}. 

The positions of all maser spots (above the level of 5$\sigma$ on the 
individual channel maps) were determined by fitting elliptical Gaussian 
models (AIPS task JMFIT). The formal errors of the fitting were 
typically 0.01--0.1\,mas in right ascension and 0.02--0.15\,mas in 
declination, depending on maser strength and structure.

\section{Results}
\subsection{12.2\,GHz emission}
In epoch 1, methanol maser emission at 12.2\,GHz towards G23.657$-$00.127 
was detected in a range from 77.41 to 86.84\,km\,s$^{-1}$, which is similar 
to the range over which 6.7\,GHz methanol maser emission was seen
(77.0--87.8\,km\,s$^{-1}$, Bartkiewicz et al. \cite{bartkiewicz}). 
In total we measured 34 maser spots in the individual channel maps. 
The strongest spot (648\,mJy\,beam$^{-1}$) appeared at $\Vlsr=82.60$\,\kms. 
The distribution of 12.2\,GHz emission closely follows the ring seen in the 
6.7\,GHz line. Indeed, all 12.2\,GHz spots have 6.7\,GHz counterparts 
but the reverse is not the case.

During the next five observations emission was seen in the same velocity 
range, but more spots were detected above the 5$\sigma$ limit on the
individual channel maps. There were 68, 86, 79, 68 
and 85 spots at epoch 2, epoch 3, epoch 4, epoch 5 and epoch 6,
respectively. The brightest spot showed a constant LSR velocity
(82.60\,km\,s$^{-1}$) but variable peak brightnesses of 619, 691, 
647, 817 and 911\,mJy\,beam$^{-1}$ at epochs 2 through 6. 
As for epoch 1, all 12.2\,GHz maser spots at all five epochs had  
6.7\,GHz counterparts. The circularly symmetric morphology of maser
emission was clearly seen in all five epochs. 
In Fig.~\ref{fig1} we present total intensity map of 12.2\,GHz emission 
from the epoch 6 data.

\subsection{Parallax measurements}
In order to determine the source parallax, we searched for spots which were 
detected in all six epochs, i.e., coinciding in velocity and position
relative to the other maser spots.
We found 19 such spots and modeled their change in position over the 
six epochs as due to the effects of annual parallax and proper motions
 (see the example in Fig.~\ref{fig2}). 
Since relative position measurements are usually dominated by systematic
errors, owing to uncompensated atmospheric delays,
formal position errors are unrealistically small leading to
rather large
$\chi^2$ per degree of freedom values ($\sim5$) for the fit. 
In order to account for systematic position errors and obtain reasonable
formal fitting uncertainties,
we added ``error floors'' of 80 and 160\,$\mu$as in right 
ascension and declination, respectively, in quadrature to the
formal fitting errors.  This resulted in $\chi^2$ per degree of freedom 
values of $\sim1$ for both the right ascension and declination data.
   
In Table~\ref{table:2} we list all 19 spots used in the fitting and
summarize the results of the parallax and proper motion fitting
individually for each sequence of spots: 
the mean parallax is 0.312\,mas, with a standard error of the mean
of 0.010\,mas (standard deviation of 0.042\,mas).
A combined fit for all spots with a single parallax, but individual proper motions
and position offsets for each spot (since all the maser spots are at the same 
distance within the measurement accuracy), gives a similar result of 
$\pi$=0.313$\pm$0.015\,mas.

In Fig.~\ref{fig3} the combined fit for the parallax with data from four 
representative spots (spots 1--4 from Table~\ref{table:2}) is presented. 
The positions of the masers after removing the proper motion and position 
offset are shown for clarity.
Combining the results of several maser spots can lead to underestimated
parallax uncertainty, since the position measurements relative to a 
background source may not be independent. Random errors 
(e.g., from the position fits due to the finite signal to noise ratios and 
possible maser spot structure) are probably not 
correlated among different maser spots.  However, systematic errors
(e.g., caused by residual atmospheric delay errors) will affect all maser spots in
one epoch in a very similar way. Thus, the formal error in the combined fit
will underestimate the true error. 

The most conservative approach in the uncertainty estimation would be to 
assume that the errors are 100\% correlated and multiply the formal error   
with $\sqrt{N}$, where $N$ 
is the number of maser spots. This would give a parallax uncertainty of 
$\pm0.015\sqrt{19}=0.065$\,mas. To estimate the effect of the systematic errors 
on our parallax measurement, we have calculated the average position 
of all maser spots in each epoch after removing their fitted
proper motions and position offsets. The average positions are shown as open
triangles in Fig.~\ref{fig3}. Such averaging should reduce the random error, 
but leave the systematic error unchanged. A parallax fit to the averaged data 
points yielded a value of 0.313$\pm$0.039\,mas, which we adopt for the 
parallax of G23.657$-$00.127. This corresponds to a distance to 
G23.657$-$00.127 of 3.19$^{+0.46}_{-0.35}$\,kpc.

The average proper motion of all spots and the standard error of the mean 
is $-$1.32$\pm$0.02 mas\,yr$^{-1}$ in 
right ascension and $-$2.96$\pm$0.03 mas\,yr$^{-1}$ in declination.
For the measured distance of 3.19~kpc, these values correspond to: 
$-20.0\pm0.3$\,\kms\ and $-44.7\pm0.5$\,\kms\ in 
right ascension and declination, respectively. All spots 
have very similar proper motions (within 0.2 mas\,yr$^{-1}$ of the average)
(Table\,\ref{table:2}). Hence, internal motions are small, as is
commonly found for 12.2\,GHz methanol masers (Moscadelli et al. \cite{moscadelli02}) and, thus,
the average proper motion closely represents the motion of the 
young star that excites the masers.

\section{Discussion}
Using our measured parallax and proper motion of the source G23.657$-$00.127, 
we can estimate its 3-dimensional motion in the Galaxy. Assuming IAU 
recommended values for the distance of the Sun from the Galactic center 
of R$_0$=8.5$\pm$0.5\,kpc and the circular rotation speed of the 
LSR of $\theta_0$=220$\pm$10 km\,s$^{-1}$,
the Hipparcos Solar Motion values (Dehnen \& Binney \cite{dehnen}),
our distance measurement of 3.19$\pm$0.40\,kpc and proper motion of 
$-$1.32$\pm$0.02\,mas\,yr$^{-1}$ in right ascension and 
$-$2.96$\pm$0.03\,mas\,yr$^{-1}$ in declination, an LSR velocity of
83$\pm$2\,km\,s$^{-1}$, we obtain a Galactocentric motion of:
\begin{eqnarray}
{\rm U'}&=&42.3  \pm  4.0\,{\rm km\,s}^{-1},\nonumber\\ 
{\rm V'}&=&222.4 \pm 11.3\,{\rm km\,s}^{-1},\nonumber\\ 
{\rm W'}&=&4.1   \pm  1.4\,{\rm km\,s}^{-1}.\nonumber
\end{eqnarray}
Here, U' denotes the radial velocity component toward  the 
Galactic Center, V' is the velocity in direction of Galactic rotation, and
W' is 
the velocity toward the North Galactic Pole. Hence, G23.657$-$00.127 is moving
with nearly the same rotation speed as the LSR and slowly out of the plane of 
the Galaxy.  However, it has a very large peculiar motion towards the Galactic 
center. This large peculiar motion explains why the source is closer than the 
near kinematic distance of $\approx5$\,kpc.

If we use a different rotation model of the Milky Way with 
R$_0$=8.0$\pm$0.5\,kpc and $\theta_0$=236$\pm$1\,km\,s$^{-1}$ which is 
consistent with the measured proper motion of the Galactic center Sgr A* 
(Reid \& Brunthaler \cite{reid}), we get: 
\begin{eqnarray}
{\rm U'}&=&33.6  \pm  3.4\,{\rm km\,s}^{-1},\nonumber\\
{\rm V'}&=&238.8 \pm  5.8\,{\rm km\,s}^{-1},\nonumber\\ 
{\rm W'}&=&4.1   \pm  1.4\,{\rm km\,s}^{-1}.\nonumber
\end{eqnarray}
Here, we find again a similar rotation speed as the LSR and a large motion 
toward the Galactic center, demonstrating that this result is not sensitive 
to the Galactic rotation model.

With a Galactocentric distance of 5.7 kpc (assuming
R$_0$=8.5$\pm$0.5\,kpc), G23.657$-$00.127 is located close to the Galactic 
bar (Benjamin et al. \cite{benjamin}). Hence, the large peculiar motion might
be caused by the gravitational potential of the central bar. Alternatively,
the large peculiar motion could be caused by an interaction between a density 
wave of a spiral arm with the molecular could. More sources in this region 
are needed to investigate this possibility in more detail.

With our new measurement of the distance, we are able to verify our earlier 
calculations of the linear size of the 6.7\,GHz ring as well as the 
bolometric luminosity using the mid- and far--infrared emission 
(Bartkiewicz et al. \cite{bartkiewicz}). Assuming the mean radius of 127\,mas
the linear size is 405\,AU. Applying the formula of Walsh et al. (\cite{walsh};
their Eq.\,(3)), we get a bolometric luminosity of 
$\leq 11.4\times10^3$L$_{\sun}$, which could be provided 
by a single B0.5 ZAMS type star (Panagia \cite{panagia}).

\section{Conclusions}
Using VLBA observations of the 12.2 GHz methanol masers in G23.657$-$00.127, we 
have determined an accurate distance of 3.19$^{+0.46}_{-0.35}$\,kpc to the 
maser ring. At this distance, the radius of the ring corresponds to a
linear size of 405 AU. We have also measured the proper motion of the source,
which we use to determine its 3-dimensional motion in the Galaxy. The source moves
with a similar rotation speed as the LSR, but shows a high peculiar motion 
of 30--40 km\,s$^{-1}$\, toward the Galactic center. This large peculiar motion
might be induced by interactions with the Galactic bar and
is responsible for the large error in the kinematic distance.

\begin{acknowledgements}
A.\,Bartkiewicz and M.\,Szymczak acknowledge support from the Polish MNiI 
grant 1P03D02729. A.\,Brunthaler was supported by the DFG Priority Programme 1177.
\end{acknowledgements}

\end{document}